# Preprint: Intuitive Evaluation of Kinect2 based Balance Measurement Software

Zhihan Lv, Vicente Penades, Sonia Blasco, Javier Chirivella, Pablo Gagliardo
FIVAN, Valencia, Spain
lvzhihan@gmail.com

## ABSTRACT
This is the preprint version of our paper on REHAB2015. A balance measurement software based on Kinect2 sensor is evaluated by comparing to golden standard balance measure platform intuitively. The software analysis the tracked body data from the user by Kinect2 sensor and get user's center of mass(CoM) as well as its motion route on a plane. The software is evaluated by several comparison tests, the evaluation results preliminarily prove the reliability of the software.

## Categories and Subject Descriptors
H.4 [**Information Systems Applications**]: Miscellaneous; D.2.8 [**Software Engineering**]: Metrics—*complexity measures, performance measures*

## General Terms
Theory

## 1. INTRODUCTION
Nowadays, there are many software and hardware approaches for human balance measurement, include various of hardware based balance measurement device [34] [15] [31] [34] [6] [4] [35] [20] [7] [2], Kinect based device [11] [12] [3] [32] [16], sensor network based device [33] [38] [18], hybrid device [30], evaluation investigation [13] [14], related interaction technology [9] [1] [10].

The evaluated software employs distinctive calculation approach of mass of center. The calculation process for each frame includes three steps. First, pre-processing step filters the noises from the range images and further removes the obstacles in the view. As shown in figure 2 up, the obstacles are marked in red color to warn the players to clean them for safety measurement, meanwhile, the obstacles are removed from the system memory so that the software can focus on calculating the valid data. In the second step, the center of mass (CoM) of the user is calculated according to the CoM of every part of user's body, as well as considering the weights of all parts. Finally, the CoM is mapped to a plane and recorded to the cloud based sever located in our headquarter. The skeleton stream of each frame is recorded and compressed, to support off-line replaying and rendering to a cartoon matchstick man, as shown in figure 2 down.

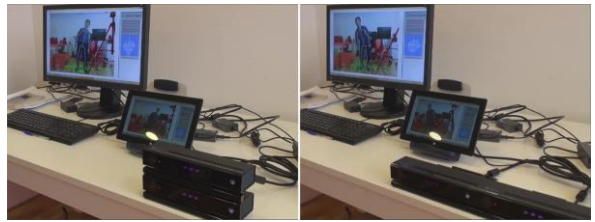

**Figure 1: Left: Place two kinect2 side by side longitudinally; Right: Place two kinect2 side by side laterally.**

The recommended minimum distance from Kinect 2 for full body tracking is 4.5ft, which is equal to 137.16 CM. In our system, we set the distance as 180-200 CM.

## 2. EVALUATION
Three comparison tests are conducted to evaluate the performance of the software.

1. The first test is to check whether the tiny offset of the kinect can affect the estimated CoM value. We compared the results of two kinects with tiny longitudinal or lateral offset, as shown in figure 1. The comparison results indicate that there isn't significant difference. Especially, the kinects with longitudinal offset got very similar balance measurement results.

2. The purpose of the second test is to evaluate how heavy can the error of estimation result reach when the kinects have far lateral offset. We compared the results of two kinects with far lateral offset and the same target point, as shown in figure 3. The results indicate the visually recognizable difference of the results from two kinects, but they are still acceptable results for our application scenarios. The results are shown as in the blue window on the right in figure 3 up.

3. In the third test, we plan to evaluate the estimated result by a known accurate reference. We compared the kinect based balance measurement software to the golden standard balance measurement platform (made by IBV), as shown in fig-ure 4. IBV platform hasn't provided SDK or recorded dynamic motion of CoM, so we use unaided eye to observe and compare the balance measurement results. The results indicate that the rough regions of the center of mass are the same. IBV platform shows more sensitive performance to capture the center of mass. Conversely, kinect based balance measurement software shows more smooth results.

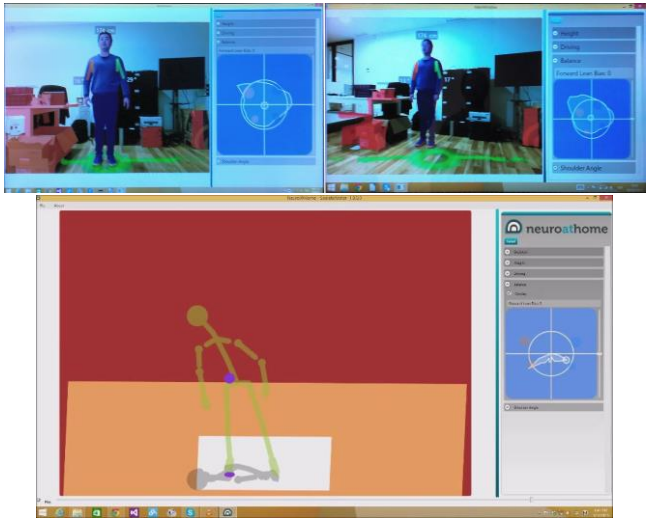

**Figure 2: Up: the real-time balance measurement comparison results from evaluation 2; Down: cartoon matchstick man visualization of offline skeleton data.**

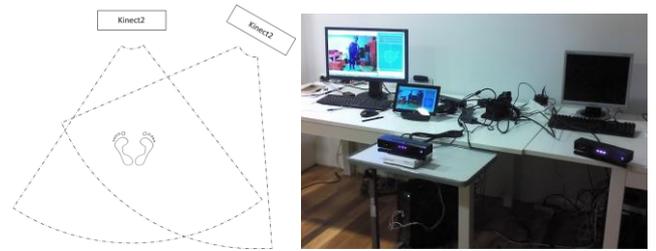

**Figure 3: Two kinects with far lateral offset and the same target point. Left: top view; Right: perspective view**

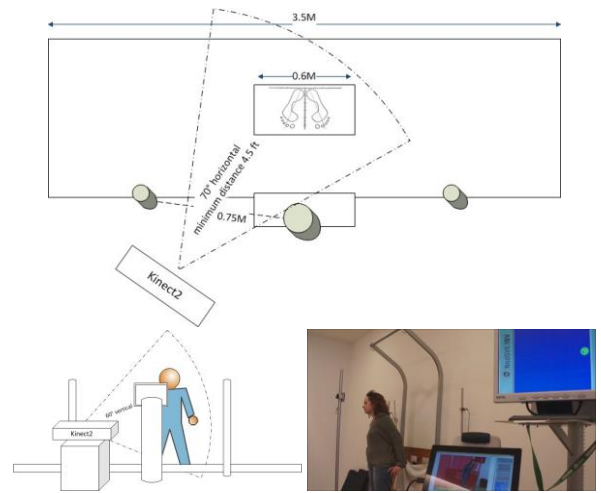

**Figure 4: Comparing the kinect based balance measurement software to the IBV platform. Up: top view; Bottom-left: frontal view; Bottom-right: perspective view**

The evaluation results proved the reliability of the demonstrated balance measurement software.

## 3. DISCUSSION

The purpose of this work is to evaluate the CoM estimation algorithm of our developed balance measurement software. The method we employed is comparing the estimated result of the software with a known golden-standard reference (IBV balance measurement platform). The evaluation proceed is fairly preliminary, since the current evaluation is based on intuitive vision feedback, but not rigorous data analysis. However, the three evaluations we have conducted are entirely able to prove that our balance measurement platform is accurate enough for the clinical observation.

During the three evaluations, some extreme conditions are tested. For example, the estimation error caused by the minimum offset and maximum offset of the location of the kinect have been evaluated. In addition, some flaws due to the limitation of kinect hardware are detected during the evaluations. The first limitation is that the accurate rate of the estimated CoM value depends on the positional relationship between the user with kinect. When the user face to the kinect sensor, the estimation result is the most accurate since the skeleton estimated from the kinect data has the most quality in this case. The second limitation is that the ground and the wall behind the user cannot be distinguished clearly from the user's body by kinect, which causes the estimation errors. To solve this practical problem, we put some white towels on the junction between the ground and the wall, as well as the the junction between the ground and the chassis of balance measurement platform, the junction between the wall and the back of the balance measurement platform.

As we mentioned before, the ultimate purposed of this research is to improve our balance measurement software. The manifested issues during the evaluations have been solved through filtering the kinect data.

## 4. CONCLUSION

The evaluations described in this paper have proved that our balance measurement software based on the CoM estimation algorithm is reliable preliminarily. The evaluations proceed also indicated some flaws of the test version. We have solved all the flaws and improved the software according to our original purpose of the evaluations. The conducted intuitive evaluation is the first step of the comprehensive evaluation of our balance measurement software, and reveals most of the existing non-core issues of the software. In next step, we plan to evaluate and further to improve the software by comparing the CoM value estimated by Wii balance board (WBB) with that by our software using numerical analysis method. Some novel technology will also be used to improve this research, e.g., Sensors [39], Virtual Rehabilitation [22] [21] [23] [28], HCI [26] [24] [25], Video Game [29] [5], Ubiquitous Computing [27], Control [42], Database [37] [40], Big Data [43] [41], Distributed Computing [36] [19] [17], Optimization Algorithm [8].

## Acknowledgment
The work is supported by LanPercept, a Marie Curie ITN funded through the 7th EU Framework Programme(316748).